\documentclass{article}
\usepackage[english]{babel}
\usepackage{theorem}
\usepackage{makeidx}
\usepackage{amsmath}
\usepackage{epic}
\usepackage{amscd}
\usepackage{psfig}
\usepackage{bbm}
\usepackage{xy}
\usepackage{brakets}
\usepackage{amssymb}
\theorembodyfont{\upshape}

\xyoption{all}
\begin{document}
\title{Towards the Evaluation of \\
       the Relevant Degrees of Freedom in \\
       Nonlinear Partial Differential Equations}
\author{Andreas Degenhard\thanks{e-mail: andreasd@icr.ac.uk,\,\,
        phone: +44 (0) 208 6613494,\,\, fax: +44 (0) 208 6610846}\\
        The Institute of Cancer Research, Department of Physics,\\
        Downs Road, Sutton, Surrey SM2 5PT, UK\\[0.4cm]
        Javier Rodr\'{\i}guez-Laguna\\
        Instituto de Matem\'aticas y F\'{\i}sica Fundamental, CSIC.\\
        C/ Serrano 123, Madrid 28006, Spain.\\[0.6cm]}
\date{}
\maketitle
\vspace{-1.0cm}
\begin{abstract}
  We investigate an operator renormalization group method to extract
  and describe the relevant degrees of freedom in the evolution of
  partial differential equations. The proposed renormalization group
  approach is formulated as an analytical method providing the
  fundamental concepts of a numerical algorithm applicable to various
  dynamical systems. We examine dynamical scaling characteristics in
  the short-time and the long-time evolution regime providing only a
  reduced number of degrees of freedom to the evolution process.
\end{abstract}
\vskip 8mm
\renewcommand{\thefootnote}{\dag}
\section{Introduction}
Based on the arguments of L.P. Kadanoff~\cite{ka66}, K.G. Wilson developed
a (numerical) renormalization group (RG) approach in the context of
critical phenomena~\cite{wi71} and later used the developed formalism to
solve the Kondo problem~\cite{wi75}. In practical RG applications to
systems with many degrees of freedom a RG transformation is established for
a stepwise elimination of the irrelevant degrees of freedom in the system.
The universality of the critical behaviour of the physical system emerges in
the RG approach in a natural way, with a set of universal exponents for each
universality class.\\
Soon after the considerable success of the RG for equilibrium systems a
generalization of the method was introduced to handle dynamic properties of
{\em spin systems}~\cite{ma76}. The goal was to obtain critical exponents for
stochastic equations and the developed generalization of the Wilson's
equilibrium theory has become known as the dynamic RG (DRG) method for
non-equilibrium systems~\cite{hh77}. The DRG method, inherently defined as a
Fourier space technique, was further developed and improved by other authors
including D. Forster, D. Nelson and M. Stephen who studied the dynamics of
the noisy Burgers equation~\cite{fns77}. In 1986 Burgers equation was
transformed into the language of growth processes by M. Kardar, G. Parisi
and Y.-C. Zhang~\cite{kpz86} later called the KPZ equation. Applying the
DRG to the KPZ equation the corresponding universal critical exponents were
derived perturbatively within various higher order
expansions~\cite{{mhkz89},{ft94}}.\\
However, the impact of dimensionality on applying the DRG method to the KPZ
or the related Burgers equation is crucial~\cite{nt92}. For dimension $d>1$
the exhibited strong coupling RG flow behaviour is not accessible by
perturbative expansions and the standard approaches fail to calculate
critical exponents~\cite{bg90}. New analytic approximations were
proposed~\cite{se92} including RG methods based on technical variations or
combining different approaches like perturbation and scaling
techniques~\cite{hh90}. Furthermore numerical simulation techniques were
developed to allow for estimating the critical exponents~\cite{{wk87},{ft90}}
although these approaches can suffer from numerical
instabilities~\cite{bc94}.\\
Contrary non-perturbative Real-Space RG techniques have been proven to
produce accurate results within strong-coupling regimes~\cite{{wh93},{so96}}
and therefore offer, in principle, the possibility to study models that
belong to the KPZ universality class. Why is it then that currently no
satisfying generally applicable Real-Space RG approach is available for
dynamical systems as this is the case for their equilibrium counterparts?
In the following we would like to answer this question in detail and we
propose a list of the fundamental problems arising in the
context.\\
Many of the modern Real-Space RG (RSRG) methods for strong-coupled systems
use an operator formalism to define the RG transformation
(RGT) \cite{{mrs96},{de00}} which under iterative application defines the
RG flow. The RGT in the operator formalism is formally based on the
application of two linear maps, a {\em truncation} or {\em fine-to-coarse}
operator and an {\em embedding} or {\em coarse-to-fine}
operator~\cite{{de00},{gmsv95}}. Usage of these two operators should be
sufficient for the construction of renormalized or effective quantities.
Within the equilibrium operator RSRG an effective observable, including in
particular the Hamiltonian, is defined by first applying the coarse-to-fine
operator followed by the original observable. Finally the fine-to-coarse
operator is used as a map back on the effective vector-space.\\
Recently N. Goldenfeld, A. McKane and Q. Hou investigated the use of
operator RSRG methods to solve partial differential equations (PDEs)
numerically~\cite{gmh98} and carried on their ideas in a further
paper~\cite{hgm01}. In their work the authors generalized the operator
concept of the RSRG to non-equilibrium system by replacing the equilibrium
Hamiltonian operator by the time evolution operator of a partial
differential equation. However the authors showed that such an approach is,
in principle, not well-defined. Furthermore the proposed approach is less
systematic and unsatisfactory for several reasons. Here we summarize the
concerns of Goldenfeld et al and complete the list by considerations of the
present authors.
\begin{enumerate}
\item Neither the coarse-to-fine operator nor the fine-to-coarse
 operator are uniquely defined. The chosen coarse-graining procedure
 does not depend on the specific dynamical system.
\item The coarse-to-fine operator and the time evolution operator do
 not commute in general. Therefore no consistent and well defined
 RSRG operator formalism can be established.
\item The geometric construction of the coarse-to-fine operator is
 unsatisfactory since it assumes the relevant degrees of freedom to
 be distributed within the long wavelength fluctuations. This is not
 necessarily the case for the evolution of many degrees of freedom
 interacting through nonlinear dynamics.
\item Instead of coarse-graining the governing PDE in a systematic
 procedure, the provided initial field configuration at time $t=0$ is
 coarse-grained with respect to a larger lattice spacing. Evolving a
 coarse-grained field configuration is not equivalent to successive
 RGTs removing the relevant degrees of freedom in time.
\item In a practical application of the operator concept introduced
 by Gol\-den\-feld et al the coarse-graining procedure and the time
 evolution procedure are successive operations. However, performing
 an equilibrium RG step followed by an evolution of the system in time
 is, in general, not equivalent to a dynamic RG step.
\item The proposed operator scheme does not allow to calculate the
 distribution of the relevant degrees of freedom in a dynamical system.
 Accordingly, RG flow equations do not yield any insight to the
 relevant physics of the system.
\end{enumerate}
In our previous work~\cite{rd01}, which can be considered as a
further development of the work of Goldenfeld et al~\cite{gmh98}, we
developed a consistent and mathematically well defined RSRG operator
approach. The basic concepts also used in this article are briefly
summarized in section \ref{concept} and generalized by including
time evolution operators. In our previous work we used the introduced
operator formalism to construct the embedding and truncation
maps by a quasi-static geometric coarse-graining. The conceptional
disadvantage of a geometric approach is that it inherently integrates
out the small scales which, because of scale interference in the
evolutionary process, are not necessarily the scales one wishes to
ignore~\cite{gmh98}. However, the technique can be easily applied to
a great variety of physical problems, even if defined on very large
lattices.\\
The essential difference in the approach proposed here
is a non-geometric construction of the coarse-to-fine and fine-to-coarse
operators involving the evolution dynamics of the particular PDE of
interest. In this work we solve all reported problems summarized above
and calculate observables to determine the universal characteristics in
a class of nonlinear PDEs. As a non-geometric generalization of our
earlier RSRG operator approach the proposed method provides a
non-perturbative real-space RG analogue to the DRG Fourier space
method.\\
We continue to present the fundamental concepts of the non-geometric
RSRG operator approach in section \ref{phys}. In particular, we develop
two concepts applicable in the short and long time regime of the
evolutionary dynamics respectively. In section \ref{ext-lin}
we solve the linear diffusion problem exactly and present an analytic
construction scheme for the coarse-to-fine and fine-to-coarse operator.
We continue in section \ref{non-lin} by generalizing the method as a
numerical algorithm applicable to nonlinear evolution equations in
general. Using transformations which are local in time allows for exact
mapping of the non-linear dynamics to linear evolution dynamics governed
by the diffusion equation.\\
In this work we denote the dependence of a function $f$ on spatial
variables $x$ and temporal variables $t$ as $f(x,t)$. Contrarily
the spatial and temporal dependence of a discretized function $f$ will
be denoted by indexing sets $i, i+1,\dots$ and $t, t+1,\dots$
respectively. Here $i, i-1$ and $i, i+1$ refer to neighbouring sites
in the lattice whereas $t-1, t, t+1$ denote successive time steps.
Not necessarily neighbouring sites are denoted as $i, j$ and by using
capital letters $I, J$ we refer to the sites of the effective lattice.
Quantities defined in this effective vector space are equipped with a
prime. A set of vectors is denoted as $\{\ket{{\bf v}_i}\}_{i=M}^N$
where the indexing set starts at $i=M$ and extends to $i=N$.
%
%%%%%%%%%%%%%%%%%%%%%%%%%%%%%%%%%%%%%%%%%%%%%%%%%%%%%
\section{The Operator Real-space RG Approach}
\label{concept}
%%%%%%%%%%%%%%%%%%%%%%%%%%%%%%%%%%%%%%%%%%%%%%%%%%%%%
%
In this work we consider examples of evolution equations of the form
\begin{align}
 \label{evoleq}
  \partial_t \phi \;=\;\tilde H \phi,
\end{align}
with $\phi=\phi(x,t)$ a function of space $x$ and time $t$ and the
operator $\tilde H$ acting as the generator of the evolution.
Although every operator $\tilde H$ acting locally in time can be
considered, we restrict our calculations for clarity to linear and
quadratic evolution operators, $H$ and $Q$ respectively,
\begin{align}
 \label{restrict}
  \tilde H \;=\; (H + Q) \;.
\end{align}
Discretizing equation (\ref{evoleq}) in space and time using
(\ref{restrict}) yields 
\begin{align}
 \label{disc_evoleq}
  {\phi}_{i, t+\Delta t} \;=\; \sum_{j=1}^N({\mathbbm{1}_{i,j}}
                    + \Delta t \,\cdot\, H_{i,j})\phi_{j, t}
                     \,+\, \Delta t \,\sum_{j,k=1}^N\, Q_{i,j,k}\phi_{j, t}\phi_{k, t}
                       \;\,:=\;f_i\left[{\bf\phi}_t\right] 
\end{align}
with $i\in\{1,\dots ,N\}$ and $N$ denoting the number of sites in
the lattice. The spatial lattice spacing is denoted as $\Delta x$
and the discrete temporal integration interval as $\Delta t$. The
function ${\bf\phi}_{t}$ defines a vector in a vector space $V^N$
and the functional $f\left[{\bf\phi}_t\right]$ defined in
(\ref{disc_evoleq}) is a map
\begin{align}
 \label{map}
   \xymatrix@1{
     f\left[{\bf\phi}_t\right] :\;\; V^{N} \;\; \ar @{^{(}->} [r] & \;\; V^{N}\;.
              }
\end{align}
Using this vector space notation the concept of a truncation or
fine-to-coarse operator $G$ together with an embedding or
coarse-to-fine operator $G^p$ is provided by the not necessarily
commuting diagram
\begin{align}
 \label{noncom_diag}
  \xymatrix{
    V^N\; \ar @{^{(}->} [d]_G \ar @{^{(}->} [r]^{f\left[{\bf\phi}_t\right]} & V^N \\
     V^M\; \ar @{^{(}->} [r]_{f'\left[{\bf\phi}'_t\right]}
       & V^M \ar @{^{(}->} [u]_{G^{p}\qquad\quad\mbox{and $\quad M\le N$} \;\;.}
  }
\end{align}
Here the truncation operator $G$ is defined by the relation
\begin{align}
 \label{trunc_def}
   {\bf\phi}'_I \;=\; \sum_{i=1}^N \, G_{I,i}\, {\bf\phi}_i \;,
\end{align}
where capital indexing letters $I\in\{1,\dots ,M\}$ refer to lattice
sites in the effective vector space $V^M$. If $M=N$ and
${\rm Ker}(G)=\emptyset$, the embedding operator $G^p$ may be chosen
as the natural inverse of the truncation operator $G$ defined as
\begin{align}
 \label{emb_def1}
   {\bf\phi}_i \;=\; \sum_{I=1}^M \, G^{p}_{i,I}\, {\bf\phi}'_I
                \;=\; \sum_{I=1}^M \, G^{-1}_{i,I}\, {\bf\phi}'_I
\end{align}
and the diagram (\ref{noncom_diag}) commutes.\\
However, within practical applications the idea is to choose $M<N$
and replace equation (\ref{disc_evoleq}) with a coarse grained
evolution equation providing less degrees of freedom. Considering
$G$ as a projector on the relevant degrees of freedom the inverse
operator $G^{-1}$ does not exist. This naturally demands for a
generalized definition of the embedding operator $G^{p}$ as the
pseudo-inverse
\begin{align}
 \label{emb_def_gen}
   {\bf\phi}_i \;=\; \sum_{I=1}^M \, G^{p}_{i,I}\, {\bf\phi'}_I
\end{align}
calculated by Singular Value Decomposition (SVD)~\cite{ptvf92}.
For $M<N$ the functional $f'\left[{\bf\phi}'_t\right]$ is defined
as in (\ref{disc_evoleq}) on an effective lattice with a reduced
number of lattice sites $M$
\begin{align}
  \xymatrix@1{
   f'\left[{\bf\phi}'_t\right] :\; V^{M} \;\; \ar @{^{(}->} [r] & \;\; V^{M}}
      \nonumber
\end{align}
\begin{align}
 \label{eff_funct}
  \mbox{and}\qquad
   \xymatrix@1{
     {\phi'}_{I, t} \;\; \ar @{|->} [r] & \;\; {\phi'}_{I, t+\Delta t}}
         \;=\; \sum_{J=1}^M & \left({\mathbbm{1}}_{I,J}
             + \Delta t \,\cdot\, H'_{I,J}\right){\phi'}_{J, t} \nonumber\\
        \, & +\, \Delta t \sum_{J,K=1}^M\, Q'_{I,J,K}{\phi'}_{J, t}{\phi'}_{K, t}\;.
\end{align}
In equation (\ref{eff_funct}) $H'$ and $Q'$ denote the effective
linear and quadratic evolution operators defined as~\cite{rd01}
\begin{align}
 \label{op_transf}
   H'_{I,J} \;:=\; G_{I,i}\,H_{i,j}\,G^{p}_{j,J}
    \qquad\mbox{and}\qquad
      Q'_{I,J,K} \;:=\; G_{I,i}\,Q_{i,j,k}\,G^{p}_{j,J}\,G^{p}_{k,K} \;\;.
\end{align}
Inserting (\ref{trunc_def}), (\ref{emb_def_gen}) and
(\ref{eff_funct}) into the non-commuting diagram (\ref{noncom_diag})
we replace equation (\ref{disc_evoleq}) by an approximate
field evolution equation on a coarse grained lattice as
\begin{align}
 \label{disc_evolapp}
  {\phi}_{i, t+\Delta t}
   &\;\approx\; \sum_{j=1}^N\sum_{J=1}^M G^{p}_{i,I}({\mathbbm{1}}_{I,J}
 + \Delta t \cdot H'_{I,J})G_{J,j} \phi_{j, t}\nonumber\\
  & \qquad\quad + \Delta t \sum_{j,k=1}^N\sum_{J,K=1}^M G^{p}_{i,I}\, Q'_{I,J,K}
          \, G_{J,j} G_{K,k}\,\phi_{j, t}\phi_{k, t}
 \nonumber\\[0.2cm]
   &\;=\; \Big[\; G^{p}\left({\mathbbm{1}}
        \, +\, \Delta t \cdot H' \, +\, \Delta t \cdot Q'\right)
                    G\,{\bf\phi}_t \;\Big]_i \quad ,
\end{align}
describing the evolution of the field ${\bf\phi}$ under a reduced
number of degrees of freedom.\\
Equation (\ref{disc_evolapp}) defines a RG transformation (RGT)
within the operator formalism and iterating the RGT defines a
RG flow. Carrying out one RGT is called a RG step (RGS) and
according to the concept developed in this section is equivalent
to an approximate field evolution using less degrees of freedom.
Equation (\ref{disc_evolapp}) defines the real-space analogue of
a RGT established within the DRG method. The field itself provides
the set of parameters used to establish a RGT~\cite{de00}.
Furthermore, equation (\ref{disc_evolapp}) fuses the
coarse-graining and the time evolution procedure and provides
a solution to the problem number 5 formulated in the introduction.
%
%%%%%%%%%%%%%%%%%%%%%%%%%%%%%%%%%%%%%%%%%%%%%%%%%%%%%
\section{Non-geometric reduction of the degrees of freedom}
\label{phys}
%%%%%%%%%%%%%%%%%%%%%%%%%%%%%%%%%%%%%%%%%%%%%%%%%%%%%
%
In this section we provide a general concept for reducing the degrees
of freedom in evolutionary systems without geometrically
coarse-graining the lattice equations. Including the temporal
characteristics of the particular partial differential equation (PDE)
into the construction of the embedding and truncation operators we
distinguish between a short-time regime and a long-time regime.
%
%%%%%%%%%%%%%%%%%%%%%%%%%%%%%%%%%%%%%%%%%%%%%%%%%%%%%
\subsection{Short-time evolution}
\label{short-phys}
%%%%%%%%%%%%%%%%%%%%%%%%%%%%%%%%%%%%%%%%%%%%%%%%%%%%%
Within the short-time regime we describe the evolution of the field
as a perturbation of the initial field configuration at time $t=0$.
Evolving the initial field ${\bf\phi}_0$ for $M$ time steps
$\Delta t$ the dynamics within this short-time interval is conserved
by the set of vectors
\begin{align}
 \label{perturb_set}
  {\cal S}\;:=\;\{ {\bf\phi}_0, {\tilde H}{\bf\phi}_0,
     {\tilde H}^2{\bf\phi}_0, \ldots, {\tilde H}^{M-1}{\bf\phi}_0 \}
\end{align}
where ${\tilde H}$ is the evolution operator defined in
(\ref{restrict}). Using the set of vectors in (\ref{perturb_set}) as
the columns and rows of the linear operators $G$ and $G^{p}$
(previously orthonormalized), these can be considered as projection
maps from and into the space $V^M$ respectively.\\
If $M\leq N$ the relation
\begin{align}
 \label{exact_evol}
   {\bf\phi}_t \;=\; G^p
      \left({\mathbbm{1}}+\Delta t \tilde H'\right)^M G\,{\bf\phi}_0
  \qquad\mbox{with}\quad \tilde H'\;=\; H'\, +\, Q'
\end{align}
governs an exact evolution of the field ${\bf\phi}_t$ on the effective
coarse-grained lattice for $t< M\Delta t$. In this case the states in
(\ref{perturb_set}) span a subspace $V^M$ of the full vector-space
$V^N$ conserving the relevant degrees of freedom for the short-time
evolutionary regime. In the considered short-time regime, we may
rewrite equation (\ref{exact_evol}) as
\begin{align}
 \label{exact_evol2}
   {\bf\phi}_t \;=\; 
         \left[\; G^p \left({\mathbbm{1}}+\Delta t \tilde H'\right) G\;\right]^M
             \,{\bf\phi}_0
\end{align}
which determines a RG flow in the short time regime and a RGS is
defined by relation (\ref{disc_evolapp}).\\
However, if $t\geq M \Delta t$ equation (\ref{exact_evol2}) is an
approximation to the evolved field ${\bf\phi}_t$. In this case the
approach is only applicable if no relevant scale interference in
the evolutionary process occurs. This is unlikely the case for
longer times in nonlinear dynamical systems and relation
(\ref{exact_evol2}) becomes a crude approximation giving rise to
numerical instabilities.\\
%
%%%%%%%%%%%%%%%%%%%%%%%%%%%%%%%%%%%%%%%%%%%%%%%%%%%%%
\subsection{Long-time evolution}
\label{long-phys}
%%%%%%%%%%%%%%%%%%%%%%%%%%%%%%%%%%%%%%%%%%%%%%%%%%%%%
Nonlinear evolution processes exhibit most of their characteristics
in the long-time regime. The asymptotic form of the field or
surface configuration in growth phenomena~\cite{kpz86} or the
formation of turbulent states out of spiral waves~\cite{ku84} are
only two examples. This gives rise to a reduction scheme for the
degrees of freedom of a dynamical system valid in the long-time
asymptotics. This in turn demands for new concepts in the
construction of the embedding and truncation operators $G^p$ and
$G$ away from any initial field configuration.\\
Therefore our task is to minimize the magnitude of the
non-commutativity in the diagram (\ref{noncom_diag}) for all times of
the evolution process. This in turn rises the important question if
such a minimum exists and if a numerical algorithm will uniquely
converge. To measure the non-commutativity in the diagram
(\ref{noncom_diag}) we calculate the difference between the
approximately evolved field using a reduced number of degrees of
freedom and the exactly evolved field as
\begin{align}
 \label{diff_field}
   {\cal E}{\bf\phi}_t \;:=\; ({\mathbbm{1}}+\Delta t \tilde H){\bf\phi}_t
              \,-\, (G^p) ({\mathbbm{1}}+\Delta t \tilde H') G\,{\bf\phi}_t \;\;.
\end{align}
We call ${\cal E}$ the error operator and generalize the operator concept
introduced in section \ref{concept} according to the commutative diagram
\begin{align}
 \label{com_diag}
  \hspace{-2.4cm}\xymatrix{
    \ar @{} [dr] |{\qquad\qquad\qquad\qquad\qquad\qquad\qquad\qquad\; =
                    \qquad\qquad\qquad\qquad\mbox{and $\quad M < N$} \;\;.}
    V^N\; \ar @{^{(}->} [d]_G \ar @{^{(}->} [r]^{f\left[{\bf\phi}_t\right]}
              & V^N\; \ar @{^{(}->} [r]^{{\cal E}{\bf\phi}_t} & V^N \\
    V^M\; \ar @{^{(}->} [r]_{f'\left[{\bf\phi}'_t\right]}
              & V^M\; \ar@/_1pc/ @{^{(}->} [ur]_{G^p}
  } \nonumber\\
\end{align}
The introduced minimization concept is one of the fundamental principles
in equilibrium operator RSRG techniques~\cite{{wh93},{gmsv95}}.
Using the established notation in the formulation of the dynamic operator
RSRG, $({\mathbbm{1}}+\Delta t H){\bf\phi}_t$ is called the target
state~\cite{wh93} and
$(G^p) ({\mathbbm{1}}+\Delta t \tilde H') G\,{\bf\phi}_t$ an optimal
representation of the target state~\cite{wh93} in the subspace
$V^M \subset V^N$.\\
We are interested in the limit ${\cal E}{\bf\phi}_t\rightarrow 0$ for
$t\gg 0$ subject to any field configuration ${\bf\phi}_0$. To exclude
the explicit field dependence from the minimization procedure we
rewrite equation (\ref{diff_field}) in operator form as
\begin{align}
 \label{diff_op}
  {\cal E} \;=\; ({\mathbbm{1}}+\Delta t H)
                   \,-\, G^p(1+\Delta t \tilde H') G \;\;.
\end{align}
To make this operator {\it as small as possible} we minimize ${\cal E}$
in the matrix notation according to the Frobenius norm
\footnote{In principle every matrix norm can be used, although
the Frobenius norm can be easily related to concepts from linear
algebra like singular value decomposition. It is defined to be the
sum of the squares of all the entries in a matrix.}.
Inserting the definitions (\ref{op_transf}) we rewrite equation
(\ref{diff_op}) using the Frobenius norm $| |_F$ as
\begin{align}
 \label{diff_op2}
  |{\cal E}|_F \; :=\; \left\{{\mathbbm{1}}- (G^p G)
      + \Delta t \Big[H - (G^p G)\,\tilde H\,(G^p G)\Big]\right\}^2 \;\;.
\end{align}
The measured magnitude of the non-commutativity in the original
diagram (\ref{noncom_diag}) has become a number $|{\cal E}|_F$
greater or equal to zero. According to equation (\ref{diff_op2})
the minimization of the error operator ${\cal E}$ only depends on
the composed operator $G^p G$. Therefore two different $G$ matrices
for which $G^p G$ is the same operator
$\xymatrix@1{V^{N} \;\; \ar @{^{(}->} [r] & \;\; V^{N}}$ yield the
same error number $|{\cal E}|_F$.\\
The matrix $G^p G$ is called the reduction operator since it governs
the field evolution under a reduced number of degrees of freedom. To
make this more obvious we rewrite $G$ and $G^p$ in terms of a
Singular Value Decomposition
\begin{align}
 \label{sing_val_dec}
  G\;=\;\sum_{i=1}^M \lambda_i \ket{{\bf u}_i}\bra{{\bf v}_i}
    \qquad\mbox{and}\qquad
      G^p\;=\;\sum_{i=1}^M \lambda^{-1}_i \ket{{\bf v}_i}\bra{{\bf u}_i} \;,
\end{align}
where $\{\ket{{\bf v}_i}\}_{i=1}^M$ and $\{\ket{{\bf u}_i}\}_{i=1}^M$
are respectively sets of $N$-dimensional and $M$-dimensional vectors.
By means of (\ref{sing_val_dec}) we define the reduction operator as
\begin{align}
 \label{reduction}
  {\cal R}\;:=\;G^p G\;=\;\sum_{i=1}^M \ket{{\bf v}_i}\bra{{\bf v}_i} \;.
\end{align}
In this notation the reduction operator ${\cal R}$ is composed of $M$
states and acts on the original vector space $V^N$, with $N > M$. Since
there are some states $\{\ket{{\bf v}_i}\}_{i=M+1}^N$ which together
with $\{{\bf v}_i\}_{i=1}^M$ represent a whole orthonormal basis of
$V_N$, the reduction operator ${\cal R}$ projects out those states
representing the less relevant degrees of freedom in the evolution
process.\\
We would like to point out that different truncation operators $G$ can
be used to construct the same reduction operator ${\cal R}= G^p G$.
In fact every RSRG approach requires a correct choice of the RG
transformation (RGT) which is not uniquely defined. In the operator
formalism the embedding and truncation operators are used to
construct the RGT~\cite{de00}. In the dynamic operator RSRG method
introduced in this work, the RGT is defined as one time step in the
field evolution equation (\ref{disc_evolapp}). From this point of
view all truncation operators yielding the same reduction operator
${\cal R}$ represent an error equivalence class as far as the
evolution error is the only concern. Other computational
considerations, such as the sparseness of the $\tilde H$ matrix
shall force us to choose one or another truncation
operator.\\
Before we discuss the field evolution according to the reduction
operator ${\cal R}$ we would like to give a remark concerning the
structure of equation (\ref{diff_op2}). Choosing $\Delta t =0$, i.e.
no dynamics is involved, equation (\ref{diff_op2}) reduces to the
geometric operator real space RG approach~\cite{rd01}. For
$\Delta t >0$ the third summand on the right hand side includes the
evolution operator of the particular PDE into the minimization
process. Furthermore this term is weighted by the time interval
$\Delta t$. Therefore the particular evolution operator is included
into the minimization process and resulting relevant degrees of
freedom for the evolutionary dynamics can depend on the chosen time
interval.\\
To reformulate the approximate field evolution in terms of the
reduction operator we insert the definitions (\ref{op_transf}) and
(\ref{reduction}) into (\ref{disc_evolapp})
resulting in
\begin{align}
 \label{evolapp_fin}
  {\phi}_{i, t+\Delta t}
   \;=\; {\cal R}\left[{\phi}_t +\Delta t \,\cdot\,
           \right( H + Q \left) \,{\cal R} \,{\phi}_t \right] \;. 
\end{align}
Within practical calculations the process of constructing ${\cal R}$
can be decomposed according to successive dimensional reduction of the
target space
\begin{align}
 \label{dim_red}
  \xymatrix@1{
               V^{N}  \;\; \ar @{^{(}->} [r] & \;\;
               V^{N-1}\;\; \ar @{^{(}->} [r] & \;\;
               V^{N-2}\;\; \ar @{^{(}->} [r] & \;\;
               \dots  \;\; \ar @{^{(}->} [r] & \;\; V^{M} \;\;.
             }
\end{align}
Within the first dimensional reduction
$\xymatrix@1{V^{N}\;\; \ar @{^{(}->} [r] & \;\; V^{N-1}}$ we have to
minimize (\ref{diff_op2}) by adjusting the $N$ components of the
vector $\ket{{\bf v}_N}$ which defines the first order reduction
operator as
\begin{align}
 \label{first_red_op}
   {\cal R}^{1}\;:=\;\sum_{i=1}^{N-1} \ket{{\bf v}_i}\bra{{\bf v}_i}
                  \;=\; {\mathbbm{1}} -  \ket{{\bf v}_N}\bra{{\bf v}_N} \;,
\end{align}
where we used the notation of relation (\ref{reduction}). Analogously
we calculate the second order reduction operator as
\begin{align}
 \label{second_red_op}
   {\cal R}^{2}\;:=\;\sum_{i=1}^{N-2} \ket{{\bf v}_i}\bra{{\bf v}_i}
               &\;=\; {\mathbbm{1}} - \ket{{\bf v}_N}\bra{{\bf v}_N}
                                 - \ket{{\bf v}_{N-1}}\bra{{\bf v}_{N-1}}
        \nonumber \\[0.2cm]
                &\;=\; {\cal R}^{1} - {\cal P}_{\ket{{\bf v}_{N-1}}}\;.
\end{align}
In (\ref{second_red_op}) we have introduced the notation of the
projection operator ${\cal P}_{\ket{\bf v}}$ defined as the
projection map ${\cal P}_{\ket{\bf v}}=(\ket{\bf v}\bra{\bf v})$ onto
the state $\ket{\bf v}$.\\
Iterating up to the $M$th order the reduction operator is calculated as
\begin{align}
 \label{Mth_red_op}
   {\cal R}^{M} \;=\; {\cal R}^{M+1} - \ket{{\bf v}_{N+1-M}}\bra{{\bf v}_{N+1-M}}
                 \;=\; {\cal R}^{M+1} - {\cal P}_{\ket{{\bf v}_{N+1-M}}} \;,
\end{align}
with ${\cal R}^{0} = {\mathbbm{1}}$. In equation (\ref{Mth_red_op}) we
have denoted by ${\cal P}_{\ket{{\bf v}_{N+1-M}}}$ the projection operator
onto the state $\ket{{\bf v}_{N+1-M}}$ which is the target to calculate
in the $M$th minimization procedure.\\
Similar calculations can be performed for the error operator ${\cal E}$.
Using the abbreviation $\ket{\bf v} := \ket{\bf v}_N$ the error operator
for the first dimensional reduction is defined as
\begin{align}
 \label{lin_stat}
  {\cal E}^1\; & =\; {\mathbbm{1}} - ({\mathbbm{1}}-\ket{v}\bra{v})
           +\Delta t\Big[\, \tilde H - ({\mathbbm{1}}-\ket{v}\bra{v})\,\tilde H\,
              ({\mathbbm{1}}-\ket{v}\bra{v}) \,\Big] \nonumber\\[0.4cm]
          \; & =\; {\cal P}_{\ket{v}}
           + \Delta t\Big(\,\tilde H {\cal P}_{\ket{v}} + {\cal P}_{\ket{v}} \tilde H
             - \bra{v}\tilde H\ket{v} {\cal P}_{\ket{v}} \,\Big) \;\;.
\end{align}
Using the iterative scheme (\ref{Mth_red_op}) we write the minimization
procedure for the state $\ket{{\bf v}_{N+1-M}}$ as
\begin{align}
 \label{lin_stat_red}
  |{\cal E}^M|_F\; & =\; {\mathbbm{1}} - {\cal R}^{M}
                  +\Delta t\Big(\,\tilde H \,
                    - \,{\cal R}^{M}\,\tilde H\, {\cal R}^{M} \,\Big) \;.
\end{align}
According to relation (\ref{lin_stat_red}) we call a truncation
$\xymatrix@1{V^{N}\;\; \ar @{^{(}->} [r] & \;\; V^{N-M}}$ of the
original vector space $V^N$ an order $M$ truncation or a reduction
of order $M$ in the degrees of freedom. The ratio
\begin{align}
 \label{red_fac}
    \lambda \; :=\; N/(N-M)
\end{align}
is defined as the reduction factor $\lambda$.
%
%%%%%%%%%%%%%%%%%%%%%%%%%%%%%%%%%%%%%%%%%%%%%%%%%%%%%
\section{An Exactly Solvable Lattice Model}
\label{ext-lin}
\begin{figure}[ht]
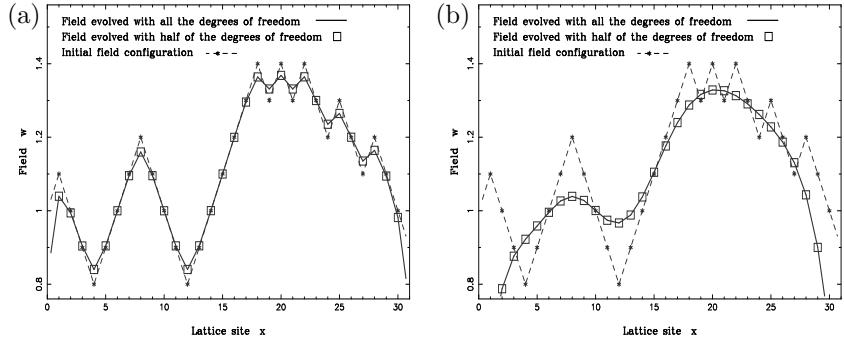

\centerline{
  \psfig{figure=short_Diff_NX=32_t=100_red16.ps,width=5.2cm,height=4.4cm}
   \hspace{0.3cm}
  \psfig{figure=short_Diff_NX=32_t=1000_red16.ps,width=5.2cm,height=4.4cm}\\[0.6cm]
    \begin{picture}(0,0)(0,0)
       \thicklines \put(-320,118){(a)}
       \thicklines \put(-158,118){(b)}
    \end{picture}
}\vspace{-0.6cm}
\caption{The initial and the evolved fields for a lattice composed
         of 32 sites after 100 and 1000 time steps (a) and (b)
         respectively. The initial field configuration is drawn by
         the dashed curve. The fields were evolved using a time
         interval $\Delta t = 1$ and a diffusion parameter
         $\nu =0.01$. The numerical minimization was performed
         up to the 16th order reduction in the degrees of freedom
         equivalent to a reduction factor of $\lambda =2$.}
\label{short_field_Diff}
\end{figure}

\begin{figure}[ht]
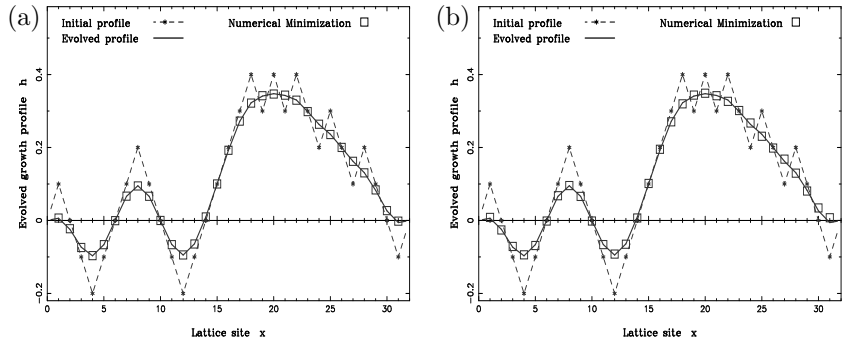

\centerline{
  \psfig{figure=field_Diff_NX=32_t=100_red8.ps,width=5.2cm,height=4.4cm}
   \hspace{0.3cm}
  \psfig{figure=field_Diff_NX=32_t=100_red16.ps,width=5.2cm,height=4.4cm}\\[0.6cm]
    \begin{picture}(0,0)(0,0)
       \thicklines \put(-320,118){(a)}
       \thicklines \put(-158,118){(b)}
    \end{picture}
}\vspace{-0.6cm}
\caption{The initial and the evolved fields for a lattice composed
         of 32 sites. The approximate field evolution ($\square$)
         is calculated according to the relation (\ref{evolapp_fin})
         with a reduction of order $M=8$ (a) and $M=16$ (b) in the
         degrees of freedom. The field is plotted after 100 time
         steps using a temporal integration  interval
         $\Delta t = 1.0$. The fields were evolved using a
         diffusion parameter $\nu =0.01$.}
\label{field_Diff_mini}
\end{figure}

%%%%%%%%%%%%%%%%%%%%%%%%%%%%%%%%%%%%%%%%%%%%%%%%%%%%%
%
In this section we treat the special case $\tilde H =H$, i.e. a
linear evolution dynamics. For both a selfadjoint and a
non-selfadjoint evolution operator the reduction operator can be
calculated exactly up to an arbitrary order in dimensional reduction
of the original vector space $V^N$. As an example for a selfadjoint
linear evolution operator we examine the diffusion equation
given by
\begin{align}
 \label{Diff_def}
  \frac{\partial {\bf w}(x,t)}{\partial t}
     \;=\;\nu\cdot{\nabla}^2 {\bf w}(x,t)\;,
\end{align}
where the diffusion coefficient $\nu$ describes the strength of the
relaxation process.\\
In figure \ref{short_field_Diff} we evolved the initial field
configuration (dashed line) according to the method described in
section \ref{short-phys} for a total time $t=100$ and $t=1000$.
% %%%%%%%%%%%%%%%%%%%%%%%%%%%%%%%%%%%%%%%%%%%%%%%%%%%%%%%%%%%%%%%%%
%
% Comparing the short-time field evolution for Diffusion.
%
% %%%%%%%%%%%%%%%%%%%%%%%%%%%%%%%%%%%%%%%%%%%%%%%%%%%%%%%%%%%%%%%%%
% %%%%%%%%%%%%%%%%%%%%%%%%%%%%%%%%%%%%%%%%%%%%%%%%%%%%%%%%%%%%%%%%%
The solid curve shows the field evolution on the original lattice
composed of 32 sites. The field evolved with half of the degrees
of freedom corresponding to an evolution on a lattice composed of
16 sites is displayed by a $\,\square\,$ for every lattice site.
The embedding and truncation operators $G^p$ and $G$ were composed
according to (\ref{perturb_set}) with M=16. For the numerical
evolution of equation (\ref{Diff_def}) we consider the following
discrete differencing of (\ref{Diff_def})
\begin{align}
 \label{Diff_def-disc}
  \frac{{\bf w}_{i,t+\Delta t}\, -\,{\bf w}_{i,t}}{\Delta t}
     \;=\;\nu\cdot\left[\,\frac{{\bf w}_{i+1,t}\,
           -\,2{\bf w}_{i,t}\, -\,{\bf w}_{i-1,t}}{\left({\Delta t}\right)^2}
         \,\right] \;.
\end{align}
In equation (\ref{Diff_def-disc}) we used a forward Euler scheme in
time~\cite{ptvf92} and the linear evolution operator $H$ is defined as
\begin{align}
 \label{Diff_H}
    H_{i,j}\; :=\; \left(\nu\cdot{\nabla}^2\right)_{i,j}\; :=\;\begin{cases}
                 \,   -2\nu/ {\left({\Delta t}\right)^2} &\quad\mbox{if}\;i=j\\
                 \quad \nu/ {\left({\Delta t}\right)^2} 
                   &\quad\mbox{if}\;j\; \mbox{nearest neighbour site of}\; i\\
                 \quad\; 0 &\quad\mbox{else} \;.
                  \end{cases}
\end{align}
From figure \ref{short_field_Diff} it is obvious that the evolution
under a reduced number of degrees of freedom does not differ from the
evolution of the field using all degrees of freedom. Even for longer
evolution times which corresponds to equation (\ref{exact_evol2})
with $M\gg N$ the field evolved with half of the degrees of freedom
does not differ from the exact evolution. This indicates
that the relevant degrees of freedom can already be deduced from the
early time dynamics. Therefore the degrees of freedom do not mix in
linear evolution, i.e. small scale dynamics and large scale dynamics
do not interfere.\\
To calculate the long-time evolution characteristics we have to
minimize $|{\cal E}|_F$ in (\ref{diff_op2}) by iterative application
of (\ref{lin_stat_red}) according to (\ref{Mth_red_op}).
We diagonalize the self-adjoint operator $H$ according to the 
orthonormal basis $\{\ket{u_i}\}_{i=1}^N$ with real eigenvalues
$E_i$ given by
\begin{align}
 \label{self_adj_eig}
   H\;=\;\sum_{i=1}^N E_i \ket{u_i}\bra{u_i}\;.
\end{align}
The eigenvalues $E_i$ are supposed to be real and in increasing 
order: $E_1\leq E_2 \leq \ldots \leq E_N$. If the basis
$\{\ket{u_i}\}_{i=1}^N$ is orthonormal the vector $\ket{v}$ in
(\ref{lin_stat}) can be decomposed according to
\begin{align}
 \label{self_adj_dec}
   \ket{v} \;=\; \sum_{j=1}^N \mu_j \ket{u_j} \;,
\end{align}
where we are assuming $\mu_i\in{\mathbbm{R}}$ for all $i$.
Using relation (\ref{self_adj_dec}) the different summands in
equation (\ref{lin_stat}) can be calculated explicitly as
\begin{align}
 \label{err_summands}
  {\cal P}_{\ket{v}} & \;=\; \sum_{i,j=1}^N \mu_i\mu_j \ket{u_i}\bra{u_j} \qquad\quad
    \bra{v}H\ket{v}\;=\; \sum_{i=1}^N \mu_i^2 E_i \;\equiv\; \braket{H}
       \nonumber\\[0.2cm]
  H{\cal P}_{\ket{v}}  &\;=\; \sum_{i,j=1}^N \mu_i\mu_j E_i \ket{u_i}\bra{u_j} \qquad\quad
    {\cal P}_{\ket{v}}H \;=\; \sum_{i,j=1}^N \mu_i\mu_j E_j \ket{u_i}\bra{u_j}
\end{align}
and relation (\ref{lin_stat}) becomes
\begin{align}
 \label{err_explicit}
  {\cal E}_{ij} \;=\; \Big[\, {\mathbbm{1}} +
                 \Delta t\left( E_i+E_j-\braket{H} \right) \,\Big]\mu_i\mu_j \;.
\end{align}
According to definition (\ref{diff_op2}) we have to minimize
\begin{align}
 \label{err_explic_norm}
  |{\cal E}|_F \;=\; \sum_{i,j=1}^N \Big[\, {\mathbbm{1}} +
           \Delta t\left( E_i+E_j-\braket{H} \right) \,\Big]^2\mu_i^2\mu_j^2 \;.
\end{align}
Expanding into explicit summands and retaining only terms which
are first order in $\Delta t$ this yields
\begin{align}
 \label{err_trunc_norm}
  |{\cal E}|_F \;=\; 1 \;+\; 2 \Delta t \braket{H} \; .
\end{align}
Therefore $\braket{H}$ must be minimized as a function of the
$\left\{\mu_i\right\}_i$, restricted to the condition
$\sum \mu_i^2=1$. To include the constraint in the minimization
we introduce a Lagrangian parameter $\lambda$ as
\begin{align}
 \label{lagself1}
  \braket{H} \;=\; \sum_{i=1}^N \mu_i^2 E_i
                  \,+\, \lambda \left(\sum_{i=1}^N \mu_i^2-1\right)\;.
\end{align}
Derivation with respect to each of the $\mu_i$ yields
\begin{align}
 \label{lagself2}
   \left(E_i \,+\, \lambda\right)\mu_i \;=\;0 \qquad\mbox{for all}\quad
            i\in\left\{1,\dots ,N\right\}\; .
\end{align}
Thus, for each $i$, either $\mu_i=0$ or $E_i=\lambda$. But, if all
eigenvalues $E_i$ are different the latter condition may apply to 
only one value of $i$. In order to fulfill the unit norm condition,
the $i$-th component must have value $\mu_i=1$. Therefore the state
$\ket{v}$ which minimizes the error operator is
\begin{align}
 \label{err_explic_solu}
  \mu_1 \;=\; 1 \qquad\mbox{and}\quad \mu_i
         \;=\; 0\;\;\;\mbox{if}\;\; i\not= 1 \;,
\end{align}
because it corresponds to the smallest eigenvalue of $H$. In the
case of degenerate eigenvalues, there is not a unique minimum vector,
but a minimum eigen-subspace.\\
To validate this result numerically we minimized the error
operator (\ref{lin_stat_red}) of order $M=8$ and $M=16$ by
calculating the set of states $\{\ket{\bf{v}_i}\}_{i=N-M}^{N}$
iteratively. To perform the minimization we used a sequential
quadratic programming (SQP) method~\cite{gmw81} provided by the
NAG Library~\cite{nag95}. The SQP algorithm used the 32 components
of each of the vectors in the set $\{\ket{\bf{v}_i}\}_{i=N-M}^{N}$
as adjusted variables and the necessary 32 gradient elements have
been numerically estimated within the NAG Library. Both for a
truncation of order $M=8$ and $M=16$ the SQP algorithm converged
into a unique minimum stated by the NAG Library as the optimal
solution found. The computer time required to evaluate a reduction
operator of order $M=16$ was about an hour on a SUN-SPARK Ultra 10
workstation (SUN Microsystems, Inc., Palo Alto, USA). In agreement
with the analytical scheme introduced above we always recalculated
the eigenstate corresponding to the highest eigenvalue within
machine precision.\\
In figure \ref{field_Diff_mini} we plotted the corresponding
field evolution of the diffusion equation (\ref{Diff_def}).
% %%%%%%%%%%%%%%%%%%%%%%%%%%%%%%%%%%%%%%%%%%%%%%%%%%%%%%%%%%%%%%%%%
%
% Comparing the field evolution for Diffusion due to minimization.
%
% %%%%%%%%%%%%%%%%%%%%%%%%%%%%%%%%%%%%%%%%%%%%%%%%%%%%%%%%%%%%%%%%%
% %%%%%%%%%%%%%%%%%%%%%%%%%%%%%%%%%%%%%%%%%%%%%%%%%%%%%%%%%%%%%%%%%
Using the approximate evolution equation (\ref{evolapp_fin}) together
with a reduction operator ${\cal R}$ of order $M=8$ no difference
occurs compared to the field evolution using all degrees of freedom.
However for a reduction factor $\lambda =2$ as shown in
figure \ref{field_Diff_mini} very small deviations can
be observed.\\
We would like to point out, that both the exact calculation and the
numerical results are highly dependent on linearity and adjointness of
the Hamiltonian, which account physically for the existence of normal
modes. A similar calculation for non-selfadjoint linear operators
using singular value decomposition is given in the appendix.
%
%%%%%%%%%%%%%%%%%%%%%%%%%%%%%%%%%%%%%%%%%%%%%%%%%%%%%
\section{A General Approach to Nonlinear Evolution Equations.}
\label{non-lin}
%%%%%%%%%%%%%%%%%%%%%%%%%%%%%%%%%%%%%%%%%%%%%%%%%%%%%
%
In this section we apply the theoretical concepts developed in
section \ref{concept} and section \ref{phys} to the deterministic
Kardar-Parisi-Zhang (KPZ) equation~\cite{kpz86} and the
deterministic Burgers equation~\cite{bu74}. Each equation provides
a different type of non-linearity to the numerical minimization
process and exhibits a different physical characteristics. As the
main milestone in the direction of nonlinear surface growth the
KPZ equation has been intensively investigated as the correct
interface equation governing the physics of lateral growth
phenomena~\cite{hz95}. The Burgers equation describes a
vorticity-free compressible fluid flow and the velocity field
${\bf v}(x,t)$ becomes the analogue of the (surface) height function
${\bf h}(x,t)$ in the KPZ equation~\cite{kr94}.\\
In this section we monitor the evolution of both fields
${\bf h}(x,t)$ and ${\bf v}(x,t)$ over different time scales. We
compute the scaling characteristics for the field fluctuations
and various observables are measured in the long time scaling
regime. We compare the performance of the evolution using a
reduced number of degrees of freedom with the evolution by direct
numerical integration in time. In addition, by using transformations
both equations can be mapped on the linear diffusion problem which
was exactly solved regarding the reduction operator ${\cal R}$ in
section \ref{ext-lin}. These transformations provide an alternative
exact construction of the reduction operator ${\cal R}$ for both
the KPZ and the Burgers equation.
%
%%%%%%%%%%%%%%%%%%%%%%%%%%%%%%%%%%%%%%%%%%%%%%%%%%%%%
\subsection{The relevant degrees of freedom in the KPZ equation}
\label{non-lin_KPZ}
\begin{figure}[ht]
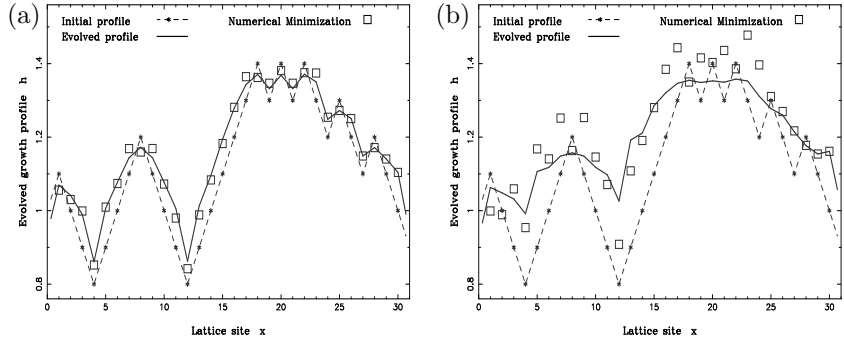

\centerline{
  \psfig{figure=short_KPZ_NX=32_t=100_red16.ps,width=5.2cm,height=4.4cm}
   \hspace{0.3cm}
  \psfig{figure=short_KPZ_NX=32_t=300_red16.ps,width=5.2cm,height=4.4cm}\\[0.6cm]
    \begin{picture}(0,0)(0,0)
       \thicklines \put(-320,118){(a)}
       \thicklines \put(-158,118){(b)}
    \end{picture}
}\vspace{-0.6cm}
\caption{The initial and the evolved surface profiles for a lattice
         composed of 32 sites after 100 time steps (a) and 300
         time steps (b) using a temporal integration interval
         $\Delta t = 1$. The dashed curve represents the initial
         surface profile and the fields were evolved using a
         surface tension $\nu =0.01$ and a growth velocity
         $\kappa =0.1$.
        }
\label{short_field_KPZ}
\end{figure}

\begin{figure}[pt]
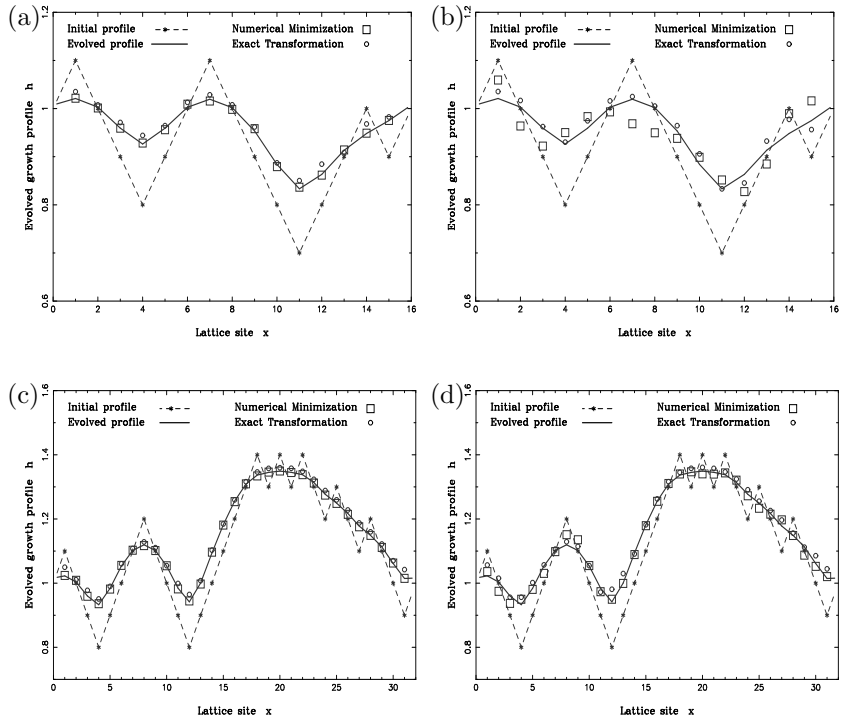

\centerline{
  \psfig{figure=field_NX=16_t=100_red=4.ps,width=5.2cm,height=4.4cm}\hspace{0.3cm}
  \psfig{figure=field_NX=16_t=100_red=8.ps,width=5.2cm,height=4.4cm}\\[0.6cm]
    \begin{picture}(0,0)(0,0)
       \thicklines \put(-320,118){(a)}
       \thicklines \put(-160,118){(b)}
    \end{picture}
}
\centerline{
  \psfig{figure=field_NX=32_t=100_red=8.ps,width=5.2cm,height=4.4cm}\hspace{0.3cm}
  \psfig{figure=field_NX=32_t=100_red=16.ps,width=5.2cm,height=4.4cm}\\[0.6cm]
    \begin{picture}(0,0)(0,0)
       \thicklines \put(-320,118){(c)}
       \thicklines \put(-160,118){(d)}
    \end{picture}
}\vspace{-0.6cm}
\caption{The initial and the evolved fields for a lattice composed
         of 16 sites ((a) and (b)) and 32 sites ((c) and (d)) after
         100 time steps using a temporal integration interval
         $\Delta t = 1.0$. The dashed curve represents the initial
         profile according to a periodic random initialization. The
         fields were evolved using a surface tension $\nu =0.01$
         and a growth velocity $\kappa =0.1$.
        }
\label{KPZfieldevol}
\end{figure}

\begin{figure}[ht]
\centerline{
  \psfig{figure=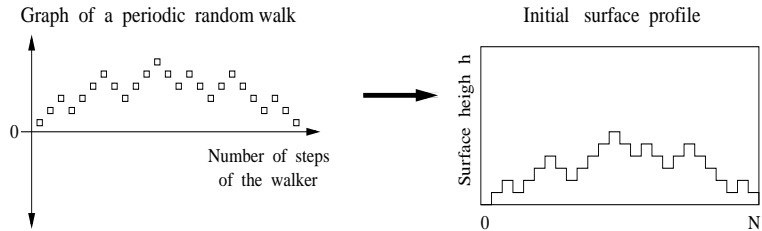,width=10.0cm,height=3.0cm}
}\vspace{-0.0cm}
\caption{An initial surface profile generated from the graph of a
         random walk using periodic boundary conditions
         (Brownian bridge).}
\label{ranwalk}
\end{figure}

\begin{figure}[pt]
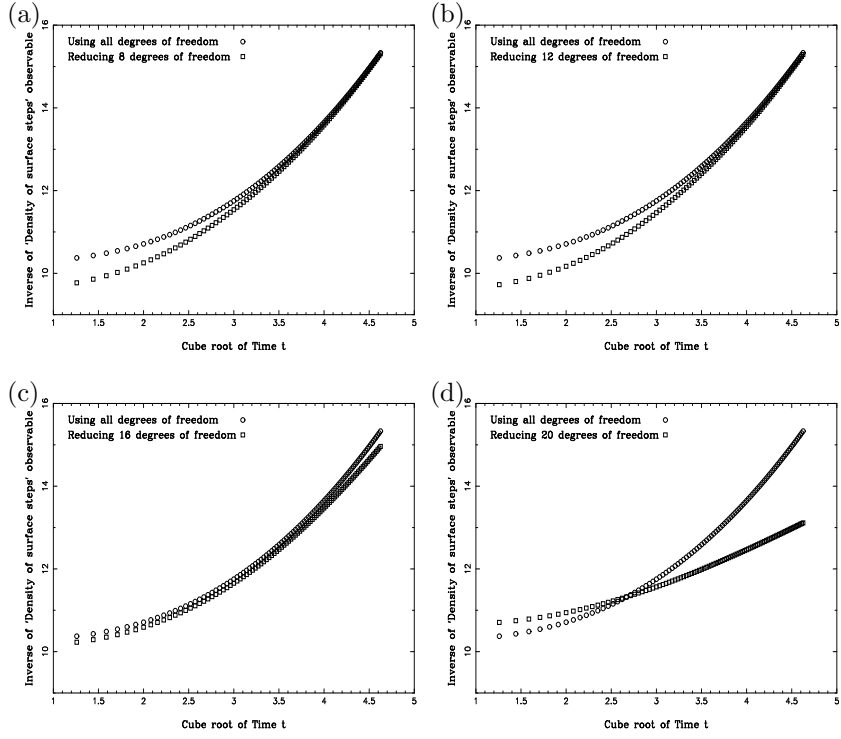

\centerline{
  \psfig{figure=dens_NX=32_t=100_red8.ps,width=5.2cm,height=4.4cm}\hspace{0.3cm}
  \psfig{figure=dens_NX=32_t=100_red12.ps,width=5.2cm,height=4.4cm}\\[0.6cm]
    \begin{picture}(0,0)(0,0)
       \thicklines \put(-320,124){(a)}
       \thicklines \put(-160,124){(b)}
    \end{picture}
}
\centerline{
  \psfig{figure=dens_NX=32_t=100_red16.ps,width=5.2cm,height=4.4cm}\hspace{0.3cm}
  \psfig{figure=dens_NX=32_t=100_red20.ps,width=5.2cm,height=4.4cm}\\[0.6cm]
    \begin{picture}(0,0)(0,0)
       \thicklines \put(-320,124){(c)}
       \thicklines \put(-160,124){(d)}
    \end{picture}
}\vspace{-0.6cm}
\caption{The inverse of the density of surface steps observable 
         plotted according to the scaling relation
         $\,\rho(t)^{-1} \sim \sqrt[3]{t}\,$ for a different
         number of eliminated degrees of freedom
         ($\lambda = 4/3, 8/5, 2$ and $8/3$). The observable
         is measured over a total time of 100 time evolution
         steps of temporal distance $\Delta t =1$ and the ensemble
         average was taken over 400 surfaces with $\zeta = 1/2$.
         The surfaces were evolved using a surface tension
         $\nu =0.01$ and a growth velocity $\kappa =0.1$.}
\label{KPZ_obs_dens}
\end{figure}

%%%%%%%%%%%%%%%%%%%%%%%%%%%%%%%%%%%%%%%%%%%%%%%%%%%%%
%
The deterministic KPZ equation is defined as
\begin{align}
 \label{KPZ_def}
  \frac{\partial {\bf h}(x,t)}{\partial t}
   \;=\; \nu {\nabla}^2 {\bf h}(x,t) + \kappa \big[\nabla {\bf h}(x,t)\big]^2\;,
\end{align}
where the first term on the right-hand side describes diffusive
relaxation of the surface and the second term introduces the desired
sideways growth. We discretize the linear part of equation
(\ref{KPZ_def}) as proposed in (\ref{Diff_def-disc}) and the
additional non-linear evolution operator defined in (\ref{restrict})
as
\begin{align}
 \label{KPZ_Q}
    Q_{i,j,k}\, :=\,
               \begin{cases}
                 \quad  \kappa/ {\left({2\Delta t}\right)^2}
                   &\;\mbox{if}\quad j=i-1;\, k=i-1\;
                                \mbox{or}\; j=i+1;\, k=i+1 \\
                 \,    -\kappa/ {\left({2\Delta t}\right)^2} 
                   &\;\mbox{if}\quad j=i-1;\, k=i+1\;
                                \mbox{or}\; j=i+1;\, k=i-1 \\
                 \qquad\; 0 &\;\mbox{else}\;.
               \end{cases}
\end{align}
In analogy to figure \ref{short_field_Diff} in section \ref{ext-lin}
figure \ref{short_field_KPZ} shows the result of the short-time RG
approach introduced in section \ref{short-phys} as applied to the
initial surface profile.
% %%%%%%%%%%%%%%%%%%%%%%%%%%%%%%%%%%%%%%%%%%%%%%%%%%%%%%%%%%%%%%%%%
%
% Comparing the short-time field evolution for the KPZ equation.
%
% %%%%%%%%%%%%%%%%%%%%%%%%%%%%%%%%%%%%%%%%%%%%%%%%%%%%%%%%%%%%%%%%%
% %%%%%%%%%%%%%%%%%%%%%%%%%%%%%%%%%%%%%%%%%%%%%%%%%%%%%%%%%%%%%%%%%
The approximately evolved field displayed in
figure \ref{short_field_KPZ} has been evolved according to a
reduction of order $M=16$ in the degrees of freedom. Although the
exact field evolution is recovered up to a time $t=100$, for $t=300$
the approximately evolved field exhibits a significant deviation from
the evolution using all the degrees of freedom. For even larger times
this results in uncontrollable numerical instabilities. The reported
behaviour is expected for RG techniques which are not able to cope
with scale interference~\cite{hgm01}.\\
To incorporate possible scale interference in the truncation procedure
it is necessary to apply the concepts developed in
section \ref{long-phys} and minimize $|{\cal E}|_F$ in (\ref{diff_op2}).
In addition to the direct numerical minimization procedure we further
propose a more sophisticated approach based on the exact solution
previously established in section \ref{ext-lin}. There it was shown that
for the linear evolution problem ${\tilde H} = H$ the dynamic RSRG
approach is equivalent to an exact diagonalization of the evolution
operator. Since equation (\ref{KPZ_def}) can be mapped to the linear
diffusion equation (\ref{Diff_def}) using the Hopf-Cole
transformation~\cite{kpz86}
\begin{align}
 \label{KPZ->Dif}
  {\bf w}(x,t) \;=\; e^{(\kappa/\nu)\cdot {\bf h}(x,t)} \;,
\end{align}
we are able to provide an exact reduction operator $R$ to evolve the
field under a reduced number of degrees of freedom
\footnote{Utilizing the Hopf-Cole transformation some authors
 reported about less instabilities in numerical integration
 schemes~\cite{{bc94},{en93}}.}.
In this case we decompose the construction of the reduction operator
into three steps. First the initial field configuration ${\bf h}(x,0)$
is transformed to ${\bf w}(x,0)$ according to equation (\ref{KPZ->Dif}).
The exactly calculated reduction operator including the relevant
degrees of freedom of the linear evolution operator $H$ is applied
within (\ref{evolapp_fin}) to evolve the field in time. Using the
inverse of transformation (\ref{KPZ->Dif}) the final field is
recovered by using a reduced number of degrees of
freedom.\\
In figure \ref{KPZfieldevol} the final evolved growth profile ${\bf h}$
after time $t=100$ is displayed for lattices of different size.
% %%%%%%%%%%%%%%%%%%%%%%%%%%%%%%%%%%%%%%%%%%%%%%%%%%%%%%%%%%%%%%%%%
%
% Comparing the field evolution for KPZ.
%
% %%%%%%%%%%%%%%%%%%%%%%%%%%%%%%%%%%%%%%%%%%%%%%%%%%%%%%%%%%%%%%%%%
% %%%%%%%%%%%%%%%%%%%%%%%%%%%%%%%%%%%%%%%%%%%%%%%%%%%%%%%%%%%%%%%%%
The approximate field evolution is performed using a reduction
operator calculated by direct numerical minimization ($square$) of
$|{\cal E}|_F$ and alternatively using the exact transformation
(\ref{KPZ->Dif}). Figure \ref{KPZfieldevol}(a) shows the evolved growth
profiles for a reduction factor of $\lambda = 4/3$ for a periodic
lattice composed of 16 sites, i.e. a truncation of order $M=4$. In
figure \ref{KPZfieldevol}(b) the initial field is equally evolved
using a reduction factor $\lambda = 2$. Compared to
figure \ref{KPZfieldevol}(a) the approximate equation
(\ref{evolapp_fin}) fails to evolve the field correctly. In
Figure \ref{KPZfieldevol}(c) and (d) the corresponding surface
evolutions are displayed for a lattice composed of 32 lattice sites.
The results displayed in figure \ref{KPZfieldevol} demonstrate that
the construction of the reduction operator ${\cal R}$ is sensitive
to finite size effects. In particular the reduction operator
constructed by direct numerical minimization on a lattice composed
of 16 sites fails to generate the characteristic paraboloid
segments~\cite{kpz86}.\\
Analogously to the numerical minimization procedure of the linear
diffusion problem carried out in section \ref{ext-lin}, the SQP
minimization algorithm was provided by the NAG Library~\cite{nag95}.
Again the optimal solution was found by the algorithm which in turn
characterizes the minimization approach in (\ref{lin_stat_red}) as a
well defined and numerically stable procedure also in the non-linear
case where no general analytical approach exists. However the
numerical minimization turns out to be much more time consuming
since the error operator norm $|{\cal E}|_F$ in equation
(\ref{diff_op2}) is increasingly complex for non-linear evolution
problems.\\
To characterize the scaling behaviour of the KPZ equation we consider
the decay of the density of surface steps for a lattice composed of
$N$ sites, defined by
\begin{align}
 \label{def_denssurf}
  \rho(t) \;:=\; \frac{1}{N}\sum^N_{i=1} \Braket{|\nabla {\bf h}_i(t)|} \;.
\end{align}
Here the brackets $\Braket{\;}$ denote averaging with respect to an
ensemble of initial surfaces ($t=0$) characterized by the covariance
\begin{align}
 \label{covar}
   \Braket{| {\bf h}_i - {\bf h}_j |}\;\sim\; | i - j |^{\zeta}
\end{align}
with the roughness exponent $\zeta$. The dynamic observable defined in
equation (\ref{def_denssurf}) obeys the scaling relation~\cite{ks88}
\begin{align}
 \label{scal_rho}
  \rho(t) \;\sim\; t^{(\zeta -1)/z}
\end{align}
where $z$ denotes the dynamic exponent for the lateral correlation
length $\xi$ determined by $\xi\sim t^{1/z}$. For the deterministic
KPZ equation a scaling relation for $z$ can be established in terms
of $\zeta$ as~\cite{ks88}
\begin{align}
 \label{scal_rel}
  z \;=\; 2 - \zeta \;.
\end{align}
To generate an ensemble of initial surfaces according to
(\ref{covar}) we initialize the surface profile by the graph of
a one dimensional Brownian bridge as visualized in figure \ref{ranwalk}.
% %%%%%%%%%%%%%%%%%%%%%%%%%%%%%%%%%%%%%%%%%%%%%%%%%%%%%%%%%%%%%%%%%
%
% Surface-initialization by a random walk.
%
% %%%%%%%%%%%%%%%%%%%%%%%%%%%%%%%%%%%%%%%%%%%%%%%%%%%%%%%%%%%%%%%%%
% %%%%%%%%%%%%%%%%%%%%%%%%%%%%%%%%%%%%%%%%%%%%%%%%%%%%%%%%%%%%%%%%%
The initial surface profile shown in figure \ref{ranwalk} corresponds to
a Brownian bridge of length $N$ introducing a finite size correction to
the correlation function which vanishes for $N\longrightarrow\infty$.
This dependence on the lattice size needs to be considered in a measurement
for the roughness exponent of the theoretical value $\zeta =1/2$~\cite{id91}.
According to the technical realization of the periodic boundary conditions
the probability for the walker to go left and right after a certain number
of surface steps depends on the remaining steps to go and the position of
the walker.\\
In figure \ref{KPZ_obs_dens}(a)-(d) the density of surface steps
observable is compared for four different reduction factors
$\lambda = 4/3, 8/5, 2$ and $8/3$.
% %%%%%%%%%%%%%%%%%%%%%%%%%%%%%%%%%%%%%%%%%%%%%%%%%%%%%%%%%%%%%%%%%
%
% 'Density of surface steps' observable for KPZ.
%
% %%%%%%%%%%%%%%%%%%%%%%%%%%%%%%%%%%%%%%%%%%%%%%%%%%%%%%%%%%%%%%%%%
% %%%%%%%%%%%%%%%%%%%%%%%%%%%%%%%%%%%%%%%%%%%%%%%%%%%%%%%%%%%%%%%%%
We expect a deviation from the exact scaling relation in the early
time regime because of the finite size dependence of the initial
surface profile constructed as a Brownian bridge. Therefore,
inserting (\ref{scal_rel}) into equation (\ref{scal_rho}) we derive
the scaling relation $\,\rho(t)^{-1} \sim t^{1/3}\,$ plotted in
figure \ref{KPZ_obs_dens} valid in the long time regime. As expected
both curves show a significant deviation from the theoretical scaling
behaviour in the short time regime. For longer times the evolution
generated by using a reduction operator providing more than half
of all the degrees of freedom for the nonlinear surface growth
process recovers the desired long-time linear behaviour. Using a
reduction factor of $\lambda =2$ or an even higher order in the
reduction of the degrees of freedom the observable measured using
the approximate field evolution displays an increasing deviation
from the measurement using a field evolved with all the degrees of
freedom.\\
As a central observation in surface growth phenomena surface
fluctuations exhibit a dynamical scaling behaviour~\cite{af93}.
The surface width of a growing surface is defined as~\cite{af93}
\begin{align}
 \label{def_width}
  {\cal W}_t \;=\; \sqrt{\frac{1}{N}\sum^N_{i=1}
                 \Big( {\bf h}_{i,t} - {\bar{h_t}}\Big)^2}
    \qquad\mbox{with}\quad 
       {\bar{h_t}}\;=\;\frac{1}{N}\sum^N_{i=1} {\bf h}_{i,t}\;.
\end{align}
In particular the scaling of the interface width (\ref{def_width})
on the length scale $N$ is expected to be of the form~\cite{hz95}
\begin{align}
 \label{scal_width}
  {\cal W}_t \;=\; N^{\zeta} f_{t/N^z}\;,
\end{align}
where $f_{t/N^z}$ denotes a scaling function different from the
one in the stochastic case. Using relation (\ref{scal_rel}) we
calculate for the KPZ equation the scaling characteristics
\begin{align}
 \label{scal_width_KPZ}
  \frac{1}{\sqrt{N}} {\cal W}_t \;=\; f_{t/N^{3/2}}\;.
\end{align}
In figure \ref{KPZ_obs_width} the surface width observable is
plotted in analogy to figure \ref{KPZ_obs_dens} measured using
equally evolved surface profiles.
% %%%%%%%%%%%%%%%%%%%%%%%%%%%%%%%%%%%%%%%%%%%%%%%%%%%%%%%%%%%%%%%%%
%
% 'Surface width' observable for KPZ.
%
% %%%%%%%%%%%%%%%%%%%%%%%%%%%%%%%%%%%%%%%%%%%%%%%%%%%%%%%%%%%%%%%%%
\begin{figure}[pt]
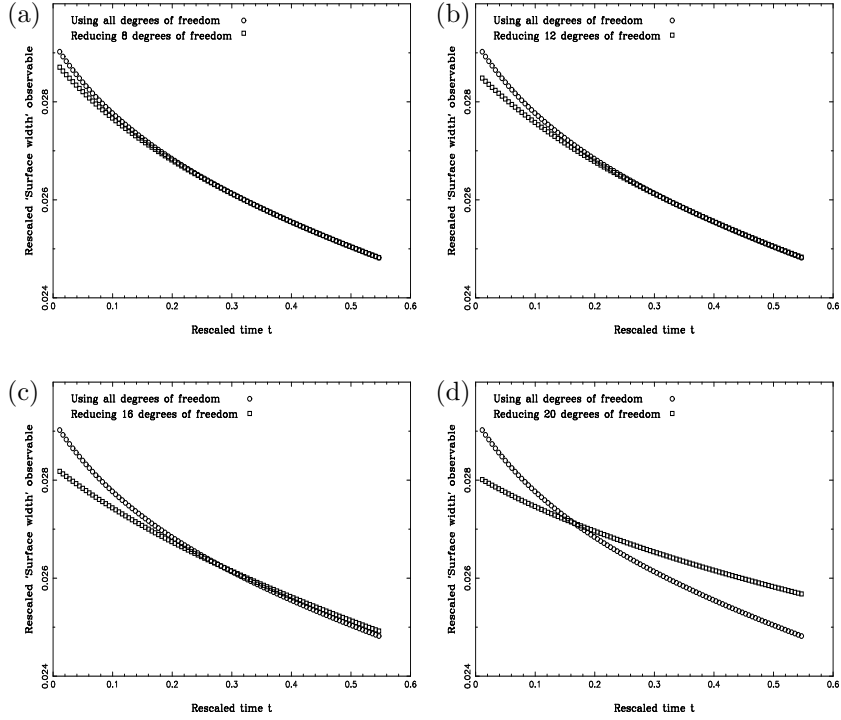

\centerline{
  \psfig{figure=width_NX=32_t=100_red8.ps,width=5.2cm,height=4.4cm}\hspace{0.3cm}
  \psfig{figure=width_NX=32_t=100_red12.ps,width=5.2cm,height=4.4cm}\\[0.6cm]
    \begin{picture}(0,0)(0,0)
       \thicklines \put(-320,118){(a)}
       \thicklines \put(-158,118){(b)}
    \end{picture}
}
\centerline{
  \psfig{figure=width_NX=32_t=100_red16.ps,width=5.2cm,height=4.4cm}\hspace{0.3cm}
  \psfig{figure=width_NX=32_t=100_red20.ps,width=5.2cm,height=4.4cm}\\[0.6cm]
    \begin{picture}(0,0)(0,0)
       \thicklines \put(-320,118){(c)}
       \thicklines \put(-158,118){(d)}
    \end{picture}
}\vspace{-0.6cm}
\caption{The surface width observable plotted according to the
         scaling characteristics of the KPZ equation. From (a)-(d)
         increasing reduction factors of $\lambda = 4/3$,
         $\lambda =8/5$, $\lambda =2$ and $\lambda =8/3$ are
         used. The observable is measured over a total time of 100
         time evolution steps of temporal distance $\Delta t =1$.
         The ensemble average was taken over 400 surfaces with
         $\zeta = 1/2$ each generated by a periodic random
         walk as displayed in figure \ref{ranwalk}. All surfaces
         were evolved using a surface tension  $\nu =0.01$ and a
         growth velocity $\kappa =0.1$.}
\label{KPZ_obs_width}
\end{figure}
% %%%%%%%%%%%%%%%%%%%%%%%%%%%%%%%%%%%%%%%%%%%%%%%%%%%%%%%%%%%%%%%%%
All measurements of the surface width observable are calculated
according to an ensemble average of 400 surface configurations to be
consistent with the literature~\cite{ks88}. The measurements for
the observable ${\cal W}$ represent the same dependence on the
reduction factor $\lambda$ as the analogue measurements for the
observable $\rho$ in figure \ref{KPZ_obs_dens}.
%
%
%%%%%%%%%%%%%%%%%%%%%%%%%%%%%%%%%%%%%%%%%%%%%%%%%%%%%
\subsection{The relevant degrees of freedom in the Burgers equation}
\label{non-lin_Burgers}
\begin{figure}[ht]
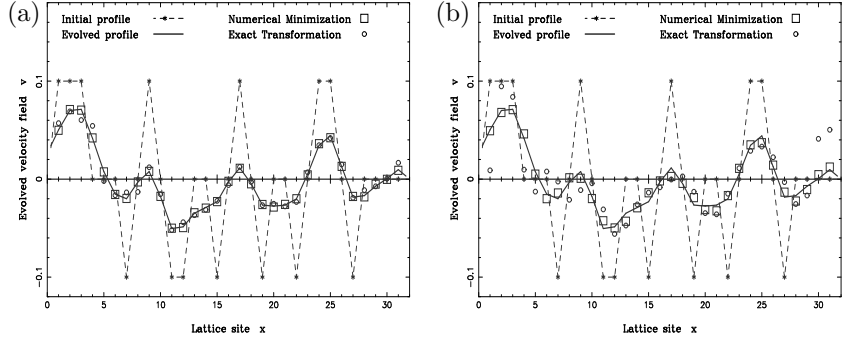

\centerline{
  \psfig{figure=field_burg_NX=32_t=100_red8.ps,width=5.2cm,height=4.4cm}
   \hspace{0.3cm}
  \psfig{figure=field_burg_NX=32_t=100_red16.ps,width=5.2cm,height=4.4cm}\\[0.6cm]
    \begin{picture}(0,0)(0,0)
       \thicklines \put(-320,118){(a)}
       \thicklines \put(-158,118){(b)}
    \end{picture}
}\vspace{-0.6cm}
\caption{The initial and the evolved velocity fields for a lattice
         composed of 32 sites after 100 time steps using a temporal
         integration interval $\Delta t = 1$. The initial velocity
         field configuration (dashed curve) is evolved with a
         reduction factor $\lambda = 4/3$ (a) and $\lambda = 2$ (b).
         The evolution using a reduced number of degrees of freedom
         is shown using a reduction operator constructed by numerical
         minimization and by using an exact transformation. The
         velocity fields were evolved with a surface tension
         $\nu =0.01$ and a growth velocity $\kappa =0.1$.
        }
\label{Burgers_fieldevol}
\end{figure}

\begin{figure}[ht]
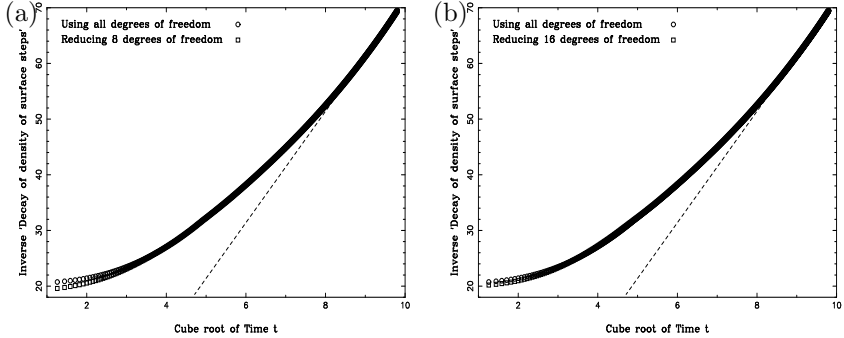

\centerline{
  \psfig{figure=dens_burg_NX=32_t=1000_red8.ps,width=5.2cm,height=4.4cm}
   \hspace{0.3cm}
  \psfig{figure=dens_burg_NX=32_t=1000_red16.ps,width=5.2cm,height=4.4cm}\\[0.6cm]
    \begin{picture}(0,0)(0,0)
       \thicklines \put(-321,118){(a)}
       \thicklines \put(-159,118){(b)}
    \end{picture}
}\vspace{-0.6cm}
\caption{The inverse of the decay of the density of surface steps
         observable plotted according to the scaling
         relation (\ref{Burg_scal}) for a reduction factor
         $\lambda = 4/3$ (a) and $\lambda = 2$ (b). The observable
         is measured over a total time of 1000 time evolution
         steps of temporal distance $\Delta t =1$. The ensemble
         average was taken over 400 random surfaces. All surfaces
         were evolved using a surface tension  $\nu =0.01$ and a
         growth velocity $\kappa =0.1$.}
\label{Burgers_obs_dens}
\end{figure}

%%%%%%%%%%%%%%%%%%%%%%%%%%%%%%%%%%%%%%%%%%%%%%%%%%%%%
%
As a quasi one dimensional analogue of the Navier-Stokes equations
the Burgers equation is defined as
\begin{align}
 \label{Burgers_def}
  \frac{\partial {\bf v}(x,t)}{\partial t}
   \;=\; \nu {\nabla}^2 {\bf v}(x,t)
        \, -\, \kappa {\bf v}(x,t)\cdot\nabla{\bf v}(x,t)\;,
\end{align}
where the parameter $\nu$ denotes the viscosity of the velocity
field ${\bf v}(x,t)$. The nonlinear term on the right hand side is
a transport or convection term in which the speed of the convection
depends on the magnitude of ${\bf v}(x,t)$
\footnote{In the original equation as introduced by Burgers
 $\kappa =1$~\cite{{bu74}}.}. Compared to the KPZ equation analyzed
in the previous section the convection term provides a different
discretization scheme to the direct numerical minimization
procedure.\\
Inserting the transformation
${\bf v}(x,t)=(\partial/\partial x){\bf h}(x,t)$ into equation
(\ref{Burgers_def}) we rederive the KPZ equation defined in
(\ref{KPZ_def}). By further applying (\ref{KPZ->Dif}) results in
the linear diffusion equation for which we can calculate the
reduction operator exactly as described in section \ref{ext-lin}.
Again, following the steps outlined for the case of the KPZ
equation below (\ref{KPZ->Dif}) we are able to provide an
alternative exact reduction operator ${\cal R}$ to evolve the
velocity field under a reduced number of degrees of
freedom.\\
Figure \ref{Burgers_fieldevol} shows an evolved velocity field
after a total evolution time $t=100$ using a lattice composed of 32
sites.
% %%%%%%%%%%%%%%%%%%%%%%%%%%%%%%%%%%%%%%%%%%%%%%%%%%%%%%%%%%%%%%%%%
%
% Comparing the velocity field evolution for Burgers equation.
%
% %%%%%%%%%%%%%%%%%%%%%%%%%%%%%%%%%%%%%%%%%%%%%%%%%%%%%%%%%%%%%%%%%
% %%%%%%%%%%%%%%%%%%%%%%%%%%%%%%%%%%%%%%%%%%%%%%%%%%%%%%%%%%%%%%%%%
In figure \ref{Burgers_fieldevol} the performance between the reduction
operator ${\cal R}$ constructed by directly minimizing $|{\cal E}|_F$
and the exact one using the proposed transformations is compared for
two different reduction factors $\lambda = 4/3$ and $\lambda = 2$.
Although both methods perform reasonably well for a reduction factor
$\lambda = 4/3$ for a reduction factor $\lambda =1/2$ the direct numerical
minimization performs far superior to the evolution incorporating
the exact transformations. The stated performance of the SQP algorithm
using the NAG Library~\cite{nag95} is equal to the examined case of the
KPZ equation in section \ref{non-lin_KPZ}, so that also in this case
the numerical minimization converges in a well defined minimum subject
to equation (\ref{diff_op2}).\\
Burgers showed that the decay of the step density follows the
dynamical scaling behaviour~\cite{{bu74}}
\begin{align}
 \label{Burg_scal}
  \bra{}{\bf v}(x,t)\ket{}
   \;\sim\; t^{-\frac{1}{3}}
\end{align}
which is equivalent to the scaling relation (\ref{scal_rho})
with $\zeta =1/2$ and $z =3/2$ using the transformation
${\bf v}(x,t)=(\partial/\partial x){\bf h}(x,t)$. Here the brackets
$\Braket{\;}$ denote an ensemble average of stationary and locally
uncorrelated initial velocity field configurations ${\bf v}(x,0)$.
In figure \ref{Burgers_obs_dens} we have plotted the asymptotic time
behaviour (\ref{Burg_scal}) over a total time $t=1000$.
% %%%%%%%%%%%%%%%%%%%%%%%%%%%%%%%%%%%%%%%%%%%%%%%%%%%%%%%%%%%%%%%%%
%
% 'Decay of density of surface steps' observable for Burgers eq.
%
% %%%%%%%%%%%%%%%%%%%%%%%%%%%%%%%%%%%%%%%%%%%%%%%%%%%%%%%%%%%%%%%%%
% %%%%%%%%%%%%%%%%%%%%%%%%%%%%%%%%%%%%%%%%%%%%%%%%%%%%%%%%%%%%%%%%%
The velocity field has been evolved by a reduction of order $M = 8$ and
$M = 16$ degrees of freedom respectively, the latter corresponding to a
$\lambda = 2$ reduction. Following Burgers~\cite{bu74} we generated an
ensemble of stationary and locally uncorrelated initial velocity field
configurations ${\bf v}_{i}(x,0)$ by calculating
\begin{align}
 \label{Burg_trans}
   {\bf v}_{i}(x,0)\;=\;(\partial/\partial x) {\bf h}_{i}(x,0)
    \qquad\mbox{and}\quad i\in\{1,\dots,400\} \;.
\end{align}
In equation (\ref{Burg_trans}) ${\bf h}_{i}(x,0)$ denotes the graph
of a Brownian bridge as shown in
figure \ref{ranwalk}.\\
To confirm the results in figure \ref{Burgers_fieldevol} and
\ref{Burgers_obs_dens} quantitatively we investigate the numerical
stability within the evolution of the velocity field in dependence of
the order of the truncation in the limit of long-time evolutionary
dynamics. Using (\ref{evolapp_fin}) together with an iterative
construction of the reduction operator ${\cal R}^M$ of order $M$
according to equation (\ref{Mth_red_op}), table \ref{sum_longtime}
summarizes the mean square deviation (MSD) as measured in comparison
with the exact velocity field evolution.
% %%%%%%%%%%%%%%%%%%%%%%%%%%%%%%%%%%%%%%%%%%%%%%%%%%%%%%%%%%%%%%%%%
%
% Table showing numerical stability for long time characteristics.
%
% %%%%%%%%%%%%%%%%%%%%%%%%%%%%%%%%%%%%%%%%%%%%%%%%%%%%%%%%%%%%%%%%%
\begin{table}[ht]
 \caption{The accuracy of the approximate velocity field evolution
          using direct numerical minimization as compared to the exact
          velocity field evolution using all degrees of freedom in the
          long-time limit. The approximate velocity field evolution is
          compared for a reduction of different orders by calculating
          the mean square deviation (MSD) between the approximately and
          the exactly evolved velocity field. All reductions calculated
          refer to an original lattice composed of 32 sites.}
 \vspace{0.2cm}
 \begin{tabular}{l||l|l|l}
  \hline\hline
  \hspace*{-0.1cm} Reduction in the & Total evolution
                   & Total evolution & Total evolution \\
  \hspace*{-0.1cm} degrees of freedom & time\, $t=100$
                     & time\, $t=1000$ & time\, $t=10000$
  \\[0.1cm] \noalign{\hrule}
   of order $M=8$
   & $6.451\cdot 10^{-4}$ & $7.906\cdot 10^{-5}$ & $3.012\cdot 10^{-6}$\cr
   of order $M=12$
   & $1.866\cdot 10^{-3}$ & $1.074\cdot 10^{-4}$ & $5.737\cdot 10^{-6}$\cr
   of order $M=16$
   & $3.600\cdot 10^{-3}$ & $1.534\cdot 10^{-4}$ & $6.428\cdot 10^{-6}$\cr
   of order $M=20$
   & $8.852\cdot 10^{-3}$ & $7.267\cdot 10^{-3}$ & $9.851\cdot 10^{-4}$\cr
   of order $M=24$
   & $1.320\cdot 10^{-2}$ & $8.976\cdot 10^{-3}$ & $1.211\cdot 10^{-3}$\cr
   of order $M=28$
   & $2.319\cdot 10^{-2}$ & $1.258\cdot 10^{-2}$ & $1.473\cdot 10^{-3}$\cr
  \hline\hline
 \end{tabular}
 \label{sum_longtime}
\end{table}
% %%%%%%%%%%%%%%%%%%%%%%%%%%%%%%%%%%%%%%%%%%%%%%%%%%%%%%%%%%%%%%%%%
The convergence in time of the MSD for each order of the reduction in
the degrees of freedom illustrates the numerical stability in the
long-time scaling regime. However, for a reduction factor $\lambda > 2$
($M=16$) the accuracy of the approximately evolved velocity field in the
long-time evolution decreases by at least two orders of magnitude. This
indicates that at least half of the available degrees of freedom are
relevant for an accurate evaluation of the long-time characteristics
in the Burgers equation.
%
%%%%%%%%%%%%%%%%%%%%%%%%%%%%%%%%%%%%%%%%%%%%%%%%%%%%%
\section{Conclusions}
%%%%%%%%%%%%%%%%%%%%%%%%%%%%%%%%%%%%%%%%%%%%%%%%%%%%%
%
In this paper we applied an operator real-space RG method to
partial differential equations (PDEs) in general. The fundamental
concept is to construct a linear map, the reduction operator
${\cal R}$, projecting to a subspace of the full vector space
including the relevant degrees of freedom governing the physics of
the evolutionary dynamics. In technical terms the reduction operator
is composed as the product of the truncation and embedding operator,
which are already established objects in the equilibrium real-space
RG literature.\\
Two basic approaches are provided for the construction of the
reduction operator ${\cal R}$. In the first approach a transformation
to a linear PDE is used for which the reduction operator can be
constructed using exact or numerically fast diagonalization
techniques. If such a transformation is missing a direct numerical
minimization is applied using a sequential quadratic programming
(SQP) algorithm to construct the reduction operator. The
advantage of the first approach over the second one is dominated
by the saving of computer time depending on the particular
nonlinear problem of interest. However, again depended on the
particular PDE, it was observed in this work that numerical
instabilities in the field evolution can prove the first approach
to be useless. Although the computer calculation time
increases with the lattice size the computed reduction operator
does not depend on a particular initial field configuration and can
be stored and applied to the evolution of any other initial field
configuration.\\
In fully developed turbulence or related non-integrable dynamical
systems the dynamics and evolution of many degrees of freedom
interacts through nonlinear PDEs. Direct numerical simulations
are difficult and show a clear demand for a reduction to the
relevant degrees of freedom. Choosing the Burgers equation to
test our RG approach we have examined one of the simplest
archetypes describing a vorticity-free, compressible fluid flow.
The measured decay of the step density can be generalized to
higher order correlation functions corresponding to structure
functions in fully developed turbulence~\cite{kr94}.\\
One of the goals of the real-space RG method presented in this
paper is certainly the reduction in complexity of the problem
yielding to a reduced amount of computer time within simulations,
after a reduction operator ${\cal R}$ has been provided. Since
this operator contains information on the relevant degrees of
freedom the proposed real-space RG method gives rise to
improved understanding of the physics of non-linear
evolutionary dynamics.\\[0.6cm]
%
%%%%%%%%%%%%%%%%%%%%%%%%%%%%%%%%%%%%%%%%%%%%%%%%%%%%%
{\leftline{\large\bf Acknowledgment}}\\[0.2cm]
%%%%%%%%%%%%%%%%%%%%%%%%%%%%%%%%%%%%%%%%%%%%%%%%%%%%%
The authors wish to thank the Department of Mathematical
Physics at the University of Bielefeld, Germany, for the
warm and kind atmosphere during their visit in which the
foundations of this work were developed. Special thanks
are given to Silvia N. Santalla for her continuous help
in the realization of the project.\\[0.8cm]
%
%%%%%%%%%%%%%%%%%%%%%%%%%%%%%%%%%%%%%%%%%%%%%%%%%%%%%
\centerline{\large\bf Appendix}\\[0.2cm]
%%%%%%%%%%%%%%%%%%%%%%%%%%%%%%%%%%%%%%%%%%%%%%%%%%%%%
In the case of a non-selfadjoint Hamiltonian, the approach
in section \ref{ext-lin} must be slightly changed. There
exists no decomposition of the functional space into
eigenvectors of the evolution operator $H$, so the natural
approach requires a singular value decomposition~\cite{ptvf92}
\begin{align}
 \label{nonself_1}
  H \;=\; \sum_{i=1}^N E_i \ket{u_i}\bra{v_i} \;.
\end{align}
In equation (\ref{nonself_1}) the vectors $\{\ket{v_i}\}_i$
denote the ``input states'' and the vectors $\{\ket{u_i}\}_i$
make up the ``output states''. It is important to state that
both sets of vectors belong to the same functional
space.\\
Changing slightly our notation for clarity $\ket{\phi}$
denotes the target state, whose removal minimizes the
error operator
\begin{align}
 \label{nonself_2}
  {\cal E} \;:=\; {\cal P}_{\ket{\phi}}
           + \Delta t\left[ H {\cal P}_{\ket{\phi}} + {\cal P}_{\ket{\phi}} H
             - \bra{\phi}H\ket{\phi} {\cal P}_{\ket{\phi}} \right]\;.
\end{align}
We decompose $\ket{\phi}$ in both the input and the
output basis as
\begin{align}
 \label{nonself_3}
  \ket{\phi}\;=\;\sum_{i=1}^N \mu_i \ket{v_i} \qquad\mbox{and}\qquad
     \ket{\phi}\;=\;\sum_{i=1}^N \chi_i \ket{u_i} \;\;.
\end{align}
Analogously to (\ref{err_summands}) we arrange the resulting
operators in (\ref{nonself_2}) using the two alternative
representations in (\ref{nonself_3}) by
\begin{align}
 \label{nonself_sums}
   {\cal P}_{\ket{\phi}} %\;=\;\ket{\phi}\bra{\phi}
     & \;=\; \sum_{i,j=1}^N \chi_i\mu_j \ket{u_i}\bra{v_j} \qquad\quad
      \bra{\phi}H\ket{\phi}
      \;=\; \sum_{i=1}^N \mu_i\chi_i E_i \;\equiv\; \braket{H} \nonumber\\[0.2cm]
   H{\cal P}_{\ket{\phi}}
       & \;=\; \sum_{i,j=1}^N E_i\mu_i\mu_j \ket{u_i}\bra{v_j} \qquad\quad
      {\cal P}_{\ket{\phi}} H
        \;=\; \sum_{i,j=1}^N \chi_i \chi_j E_j \ket{u_i}\bra{v_j} \quad .
\end{align}
Therefore, in the basis of the evolution operator
\begin{align}
 \label{nonself_errlin}
   {\cal E} & \;=\; \sum_{i,j=1}^N {\cal E}_{ij}\ket{u_i}\bra{v_j}
       \nonumber\\[0.2cm] \quad\mbox{with}\qquad
     {\cal E}_{ij}\;=\; \chi_i\mu_j & + \Delta t \Big( E_i\mu_i\mu_j
                  + \chi_i\chi_jE_j - \braket{H}\chi_i\mu_j \Big) \;.
\end{align}
Squaring and retaining only terms which are linear in $\Delta t$
\begin{align}
 \label{nonself_errsqu}
  {\cal E}_{ij}^2 \;\approx\; \chi_i^2\mu_i^2
           + 2\Delta t\Big( E_i\mu_i\chi_i\mu_j^2 + E_j\chi_i^2\chi_j\mu_j
            - \braket{H} \chi_i^2\mu_j^2 \Big) \;.
\end{align}
Summation over all the values in (\ref{nonself_errsqu}) yields
\begin{align}
 \label{nonself_errsqu2}
   \sum_{i,j=1}^N {\cal E}_{ij}^2 \;=\; 1\, +\,  2\cdot\Delta t \braket{H}
\end{align}
to first order in $\Delta t$. The result is equivalent to equation
(\ref{err_trunc_norm}) and a again $\braket{H}$ must be minimized.
However one has to remember that the set of numbers $\{\mu_i\}_i$
and $\{\chi_i\}_i$ are not independent since they represent the
{\sl same} field.\\
The idea is to represent the vectors $\{\ket{v_i}\}_i$ in the basis
of the vectors $\{\ket{u_i}\}_i$ according to
\begin{align}
 \label{nonself_bas1}
  \ket{v_i} \;=\; \sum_{j=1}^N C_{ij}\ket{u_j}
\end{align}
The matrix $C_{ij}$ may be obtained by the SVD of the evolution
operator $H$ using
\begin{align}
 \label{nonself_bas2}
   C_{ij} \;\equiv\; \ket{v_j}\bra{u_i} \;.
\end{align}
The equality of both representations in (\ref{nonself_3}) yields a
relation between the $\{\mu_i\}_i$ and the $\{\chi_i\}_i$:\\
\begin{align}
 \label{nonself_equiv}
  \sum_{i=1}^N \mu_i\ket{v_i}
   \,=\, \sum_{i,j=1}^N \mu_i C_{ij} \ket{u_j}
    \,=\, \sum_{j=1}^N \left( \sum_{i=1}^N C_{ij}\mu_i \right) \ket{u_j}
     \;=\; \sum_{j=1}^N \chi_j \ket{u_j}
\end{align}
As the $\{\ket{u_j}\}_i$ form an orthonormal basis we may deduce
\begin{align}
 \label{nonself_dedu}
   \chi_j \;=\; \sum_{i=1}^N C_{ij}\mu_i \;.
\end{align}
Using (\ref{nonself_sums}) and inserting (\ref{nonself_dedu})
the value to be minimized equals
\begin{align}
 \label{nonself_finmin}
  \braket{H} \;=\; \bra{\phi}H\ket{\phi}
              \;=\; \sum_{i=1}^N \mu_i\chi_i E_i
               \;=\; \sum_{i,j=1}^N E_i \mu_i C_{ji}\mu_j \;\;.
\end{align}
Equation (\ref{nonself_finmin}) defines a quadratic form based
on a matrix which is not symmetric. Defining
\begin{align}
 \label{nonself_symm}
   K_{ij}\equiv\frac{1}{2}\left( E_iC_{ji}+ E_jC_{ij} \right)
\end{align}
the problem is reduced to the minimization of the quadratic form
\begin{align}
 \label{nonself_finquadf}
  \braket{H}=\sum_{i,j=1}^N K_{ij} \mu_i\mu_j
\end{align}
with the restriction $\sum_i \mu_i^2=1$. This problem is equivalent
to the self-adjoint case examined in section \ref{ext-lin} and
the solution is the lowest eigenvalue of the
matrix $K$.\\
The eigenvector corresponding to the minimum eigenvalue of the
matrix $K$ is to be interpreted as a set of $\mu_i$ values.
Thus, the $\ket{\phi}$ eigenvector shall be given by the
calculation of $\sum_i \mu_i\ket{u_i}$ with the appropriate
values of $\mu_i$.\\[0.8cm]
%
%
%%%%%%%%%%%%%%%%%%%%%%%%%%%%%%%%%%%%%%%%%%%%%%%%%%%%%
%

\end{document}